\documentclass[10pt]{iopart}

\usepackage{graphicx}
\usepackage{subfigure}
\usepackage{verbatim}
\usepackage{dcolumn}
\usepackage{bm}
\usepackage{epsf}
\usepackage{color}
\usepackage[colorlinks=true,citecolor=blue,linkcolor=blue,urlcolor=blue]{hyperref}
\usepackage{hhline}
\usepackage{amssymb}
\usepackage{iopams}
\usepackage{cite}
\usepackage{tikz}
\usetikzlibrary{arrows,shapes,calc,matrix}

  \expandafter\let\csname equation*\endcsname\relax
  \expandafter\let\csname endequation*\endcsname\relax
\usepackage{amsmath}

\makeatletter
\newcommand\footnoteref[1]{\protected@xdef\@thefnmark{\ref{#1}}\@footnotemark}
\makeatother

\newcommand{\bra}[1]{\left\langle #1\right|}
\newcommand{\ket}[1]{\left|#1\right\rangle}

\newcommand{\id}{\mathbb{I}}

\newcommand{\bla}{bla\\bla\\bla\\bla\\bla}

\newcommand{\mc}[1]{\mathcal{#1}}

\newcommand{\draftmode}{1}    
\newcommand{\notetoself}[1]{\ifnum \draftmode=1 {\color[rgb]{0,0,0.8} [#1]} \fi}  
\newcommand{\cuttext}[1]{\ifnum \draftmode=1 {\color[rgb]{0,0.5,0} [#1]} \fi}  
\newcommand{\warntext}[1]{\ifnum \draftmode=1 {\color[rgb]{0.9,0.6,0} #1} \else {#1} \color{black} \fi}

\begin{document}

\title{Demonstration of \emph{entanglement assisted invariance} on IBM's Quantum Experience}

\author{Sebastian Deffner}
\address{Department of Physics, University of Maryland Baltimore County, Baltimore, MD 21250, USA}

\date{}

\begin{abstract}
Quantum entanglement is among the most fundamental, yet from classical intuition also most surprising properties of the fully quantum nature of physical reality. We report several experiments performed on IBM's Quantum Experience demonstrating envariance -- entanglement assisted invariance. Envariance  is a recently discovered symmetry of composite quantum systems, which is at the foundational origin of physics and a quantum phenomenon of pure states. These very easily reproducible and freely accessible experiments on Quantum Experience provide simple tools to study the properties of envariance, and we illustrate this for several cases with ``quantum universes'' consisting of up to five qubits.
\end{abstract}

\section{Entanglement -- at the foundational origin of physics}

In many aspects classical intuition fails to describe the fully quantum nature of physical reality. In particular, the way quantum states share information and how they are correlated can at times be a bit spooky \cite{Born1971}. However, nowadays there is little doubt that \emph{quantum entanglement} is real \cite{Shalm2015,Hensen2015}, and that the quantum mechanical description of physical reality is, indeed, complete \cite{Einstein1935,Schrodinger1935,Schrodinger1936}. 

Although quantum entanglement might be counterintuitive from a classical point of view, its very nature is at the core of the foundations of physics. For instance, it has been well-established that statistical mechanics is deeply rooted in properties and consequences of entangled quantum states \cite{Goldstein2006,Popescu2006,Deffner2015,Deffner2016,Chiribella2016}, and quantum entanglement is essentially what will make potential quantum computers work \cite{Ladd2010}.

An important and at the same time particularly peculiar consequence of entanglement is the existence of purely non-classical symmetries. Consider a quantum system, $\mc{S}$, which is fully entangled with an environment, $\mc{E}$, and let $\ket{\psi_\mc{SE}}$ denote the composite state in a ``quantum universe'', $\mc{S}\otimes\mc{E}$. Then $\ket{\psi_\mc{SE}}$ is called \emph{envariant} under a unitary map $U_\mc{S}=u_\mc{S}\otimes\id_\mc{E}$, if $\ket{\psi_\mc{SE}}$ exhibits entanglement assisted invariance under $U_\mc{S}$. By this we mean that $\ket{\psi_\mc{SE}}$ is envariant under $U_\mc{S}$, if there exists another unitary $U_\mc{E}=\id_\mc{S}\otimes u_\mc{E}$ which only acts on $\mc{E}$ but not on $\mc{S}$ such that,
\begin{equation}
\label{eq01}
\begin{split}
U_\mc{S} \ket{\psi_\mc{SE}} &=\left(u_\mc{S}\otimes\id_\mc{E}\right)\ket{\psi_\mc{SE}}=\ket{\eta_\mc{SE}}\\
U_\mc{E}\ket{\eta_\mc{SE}}&=\left(\id_\mc{S}\otimes u_\mc{E}\right)\ket{\eta_\mc{SE}}=\ket{\psi_\mc{SE}}\,.
\end{split}
\end{equation}
Thus, $U_\mc{E}$ ``does the job'' of the inverse map of $U_\mc{S}$ on $\mc{S}$ -- assisted by the quantum environment $\mc{E}$. This quantum symmetry, aka \emph{envariance} \cite{Zurek2003a} is a symmetry of pure quantum states, which has no direct classical analog.\emph{Pure states} of classical, composite systems are given by Cartesian, rather than by tensor products, and hence such an environment assisted inverse cannot exist \cite{Zurek2005}.

The importance of envariance for the foundations of physics can hardly be underestimated. Originally discovered in a derivation of Born's rule \cite{Zurek2003a,Zurek2005,Schlosshauer2005,Zurek2011}, it quickly became clear that the emergence of classical reality from quantum physics is deeply rooted in envariantly shared quantum information \cite{Zurek2009,Zurek2014,Zwolack2016}. Nevertheless, the experimental demonstration of envariance poses formidable technological challenges that were only recently overcome in two (quantum) optical experiments \cite{Vermeyden2015,Harris2016}. 

Earlier this year IBM made a fully functional, 5-qubit quantum computer publicly available via the world wide web \cite{IBM2016}. The architecture is based on 5 transmon qubits in a star geometry, and a full Clifford algebra is available \cite{IBM2016}. The system is freely available through IBM's cloud, and several experiments have already been reported, such as the verification of entropic uncertainty relations \cite{Berta2016}, the implementation of quantum error correction \cite{Takita2016,Devitt2016}, the experimental test of Mermin inequalities \cite{Alsina2016}, ``easy'' quantum state tomography \cite{Rundle2016}, and a demonstration of quantum teleportation \cite{Fedortchenko2016}.

In this paper we report a simple and easily reproducible demonstration of envariance on IBM's \emph{Quantum Experience}. Thus, the purpose of this demonstration is twofold: (i) we report a freely accessible, and pedagogical experiment demonstrating properties of entanglement with the first quantum computer in the cloud, and (ii) we demonstrate envariance in a ``universe'' consisting of up to five qubits, which goes beyond the technically challenging experiments in quantum optics, which (so far) have been restricted to only two qubits.

In the following we briefly describe the 5-qubit quantum computer in Sec.~\ref{IBM}, before we report the outcome of several experiments demonstrating envariance in Sec.~\ref{experiment}. The analysis is concluded with a few remarks in Sec.~\ref{con}.

\section{\label{IBM}IBM's Quantum Experience}

In May 2016 IBM made a universal quantum computer available to the public via the IBM cloud \cite{IBM2016,IBM2016a}, and which is housed at the IBM T.J. Watson Research Center in New York \cite{IBM2016a}. Currently, \emph{Quantum Experience} consists of 5 transmon qubits \cite{Koch2007,Schreier2008}, that are connected in a star geometry. Principally, a full Clifford algebra \cite{Nielsen2010} is available, with the exception that the CNOT gate can only be performed on the center qubit with any of the four peripheral qubits.

Programming the quantum computer is conveniently provided via a graphical user interface, which IBM dubbed the \emph{composer}. The interface is so accessible and operational that it has been suggested that now any undergraduate student could perform experiments on a real-life quantum computer \cite{Fedortchenko2016}. However, the system is robust enough to also allow research experiments \cite{Devitt2016}, and in particular the demonstration of quantum entanglement \cite{Alsina2016}. Quantum Experience is calibrated daily, and typically its temperature is around a few mK and the decoherence times of the single qubits are about 50-100$\mu$sec. 

However, it should be emphasized that a shortcoming of Quantum Experience is that only projective measurements of the single qubit states in $z$-direction qubits can be performed, i.e., it can be measured whether the single qubits are in $\ket{\uparrow}$ or $\ket{\downarrow}$. No quantum tomography for the joint state is available, which means that no \emph{direct} measurements of correlations between qubits can be performed.

\section{\label{experiment}Experimental demonstration of envariance}

As outlined above, envarience is a purely quantum symmetry of entangled quantum systems. A joint quantum state, $\ket{\psi_\mc{SE}}$ living in a ``quantum universe'', $\mc{S}\otimes\mc{E}$, is called envariant under a unitary map, $U_\mc{S}$, which acts only on the system, $\mc{S}$, if the action of $U_\mc{S}$ can be undone by another unitary acting only its complement $\mc{E}$. The principle is most easily illustrated with a simple example:

Suppose $\mc{S}$ and $\mc{E}$ are each given by two-level systems, where $\{\ket{\uparrow}_\mc{S}, \ket{\downarrow}_\mc{S}\}$ are the eigenstates of $\mc{S}$ and $\{\ket{\uparrow}_\mc{E}, \ket{\downarrow}_\mc{E}\}$ span $\mc{E}$. Now, further assume that $\ket{\psi_\mc{SE}}\propto\ket{\uparrow}_\mc{S}\otimes\ket{\uparrow}_\mc{E}+\ket{\downarrow}_\mc{S}\otimes\ket{\downarrow}_\mc{E}$ and $U_\mc{S}$ is a \textit{swap} in $\mc{S}$ -- $U_\mc{S}$ ``flips'' $\mc{S}$'s spin. Then, we have
\begin{equation}
\label{eq02}
 \begin{tikzpicture}[>=stealth,baseline,anchor=base,inner sep=0pt]
      \matrix (foil) [matrix of math nodes,nodes={minimum height=0.5em}] {
         & {\color{blue}\ket{\uparrow}_\mc{S}} & \otimes & \ket{\uparrow}_\mc{E} &  & + &  & {\color{blue}\ket{\downarrow}_\mc{S}} & \otimes & \ket{\downarrow}_\mc{E} &  \xrightarrow{\quad {\color{blue}U_\mc{S}}\quad} 
{\color{blue}\ket{\downarrow}_\mc{S}}\otimes\ket{\uparrow}_\mc{E}+{\color{blue}\ket{\uparrow}_\mc{S}}\otimes\ket{\downarrow}_\mc{E}\,.\\
      };
      \path[->] ($(foil-1-2.north)+(0,1ex)$)   edge[blue,bend left=45]    ($(foil-1-8.north)+(0,1ex)$);
      \path[<-] ($(foil-1-2.south)-(0,1ex)$)   edge[blue,bend left=-45]    ($(foil-1-8.south)-(0,1ex)$);
    \end{tikzpicture}
\end{equation}
The action of $U_\mc{S}$ on $\ket{\psi}_\mc{SE}$ can be restored by a swap, $U_\mc{E}$, on $\mc{E}$,
  \begin{equation}
\label{eq03}
    \begin{tikzpicture}[>=stealth,baseline,anchor=base,inner sep=0pt]
      \matrix (foil) [matrix of math nodes,nodes={minimum height=0.5em}] {
         & \ket{\downarrow}_\mc{S} & \otimes & {\color{red}\ket{\uparrow}_\mc{E}} &  & + &  & \ket{\uparrow}_\mc{S} & \otimes & {\color{red}\ket{\downarrow}_\mc{E}} & \xrightarrow{\quad {\color{red}U_\mc{E}}\quad} 
\ket{\downarrow}_\mc{S}\otimes{\color{red}\ket{\downarrow}_\mc{E}}+\ket{\uparrow}_\mc{S}\otimes{\color{red}\ket{\uparrow}_\mc{E}}\,. \\
      };
      \path[->]  ($(foil-1-4.north)+(0,1ex)$) edge[red,bend left=45] ($(foil-1-10.north)+(0,1ex)$);
       \path[<-]  ($(foil-1-4.south)-(0,1ex)$) edge[red,bend left=-45] ($(foil-1-10.south)-(0,1ex)$);
    \end{tikzpicture}
  \end{equation}
Thus, the swap $U_\mc{E}$ on $\mc{E}$ restores the pre-swap $\ket{\psi}_\mc{SE}$ without ``touching'' $\mc{S}$, i.e., the global state is restored by solely acting on $\mc{E}$. 

\subsection{Swaps in multiple qubits}

The above example can be readily implemented on Quantum Experience. We performed experiments with quantum ``universes'' consisting of 2, 3 and 5 qubits.

\subsubsection{Swaps in 2 qubits}

We begin with the simplest situation, in which the quantum system $\mc{S}$ is given by a single qubit, and the environment $\mc{E}$ is also only a single qubit.

\begin{figure}
\centering
\includegraphics[trim={0 1cm 28cm 3cm},clip,width=.9\textwidth]{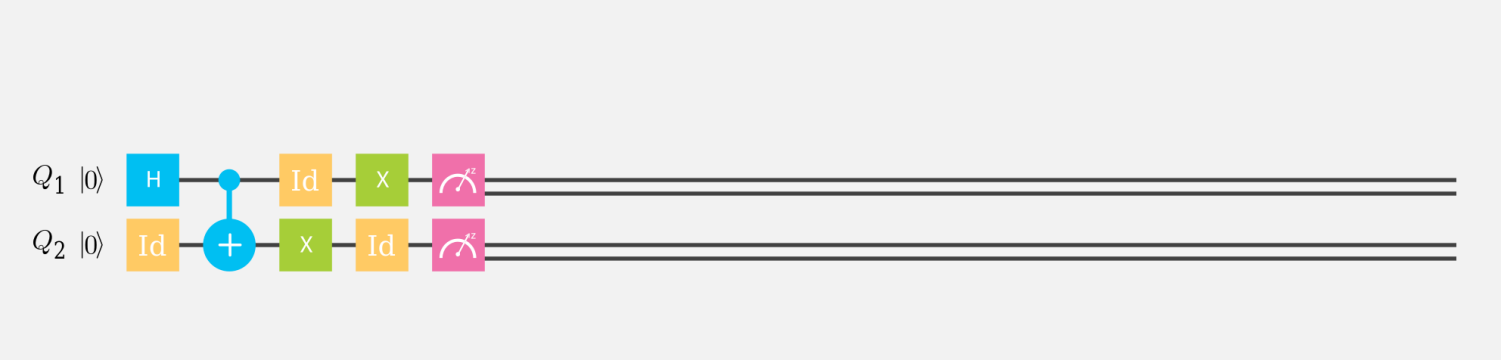}
\caption{\label{fig1} Depiction of a quantum algorithm equivalent to the envariant swap operation, cf. Eqs.~\eqref{eq02}-\eqref{eq03}. $Q2$ is chosen as system, $\mc{S}$, and $Q1$ is the quantum environment, $\mc{E}$. Per default all qubits are initially prepared in their ground states, $\ket{\downarrow}=\ket{0}$. Our desired initial state, $\ket{\psi_\mc{SE}}\propto\ket{\uparrow\uparrow}+\ket{\downarrow\downarrow}$, is obtained by performing a Hadamard-gate, $H$ (blue box) on $Q1$, followed by a CNOT. The swap in $\mc{S}$ is realized by a bitflip, i.e., a $\sigma_x$-gate ($X$ in Quantum Experience's notation, green boxes), which can be counteracted by a  bitflip in $\mc{E}$. The pink boxes at the end of the algorithm depict the projective measurements of the qubits' states.}
\end{figure}

Per default all qubits are prepared on Quantum Experience in the down state, and hence we first have to prepare $\ket{\psi_\mc{SE}}\propto\ket{\uparrow}_\mc{S}\otimes\ket{\uparrow}_\mc{E}+\ket{\downarrow}_\mc{S}\otimes\ket{\downarrow}_\mc{E}\equiv\ket{\uparrow\uparrow}+\ket{\downarrow\downarrow}$. The whole algorithm is depicted in Fig.~\ref{fig1}, which is taken from the graphical user interface of Quantum Experience. For this experiment qubit $Q2$ was chosen as system $\mc{S}$ and qubit $Q1$ was chosen as environment $\mc{E}$. 

Preparing the initial state, $\ket{\psi_\mc{SE}}$, is achieved by applying a Hadamard-gate on one qubit ($Q1$), followed by a CNOT operation entangling $\mc{S}$ and $\mc{E}$. The first swap \eqref{eq02} is then realized by performing a $\sigma_x$-gate on $\mc{S}$, which can be ``counteracted'' by another $\sigma_x$-gate on $\mc{E}$.  Finally, the states of $\mc{S}$ and $\mc{E}$ are measured separately\footnote{Note again that Quantum Experience does not provide measurements of the joint state, but that qubits can only be measured separately.}.

The algorithm was run 4 times for different peripheral qubits, with 1024 or 8192 shots, where 8192 is the maximal number of shots allowed on Quantum Experience for a ``single run''. The results of the experiments are summarized in Tab.~\ref{tab1}. 
\begin{table}
\centering
\begin{tabular}{c||c|c|c|c||c}
& $\ket{\downarrow\downarrow}$&$\ket{\downarrow\uparrow}$ &$\ket{\uparrow\downarrow}$ &$\ket{\uparrow\uparrow}$&$B$ \\
\hline\hline
theory &0.5 & 0 &0 &0.5&1\\
\hline
run 1 (1024) &0.475 & 0.046 &0.037 &0.442&0.957\\
\hline
run 2 (8192) & 0.468& 0.043&0.041 &0.448&0.957\\
\hline
run 3 (8192) &0.481 &0.042 & 0.035 & 0.442&0.961\\
\hline
run 4 (1024) & 0.435 & 0.053 & 0.057 & 0.456&0.944\\
\end{tabular}
\caption{\label{tab1} Relative frequencies of the final states obtained from the algorithm depicted in Fig.~\ref{fig1} for several separate runs with either 1024 or 8192 shots;  $B$  is the classical fidelity \eqref{eq04} with respect to the theoretically expected values.}
\end{table}
To assess the quality of the experiment we compute the classical fidelity, i.e., the Bhattacharyya coefficient \footnote{Quantum Experience only provides  projective measurements on single spins. Thus, full quantum tomograhpy of the joint multi-qubit state is neither partical nor readily available.}, which is given by \cite{Bhattacharyya1946}
\begin{equation}
\label{eq04}
B=\sum_i\sqrt{p_i q_i}\,,
\end{equation}
where $p_i$ is the frequency of the observed states, $\{\ket{\downarrow\downarrow}, \ket{\downarrow\uparrow}, \ket{\uparrow\downarrow}, \ket{\downarrow\downarrow}\}$, and $q_i$ is the theoretical prediction $q_{\downarrow\downarrow}=0.5$,  $q_{\downarrow\uparrow}=0$, $q_{\uparrow\downarrow}=0$, and $q_{\uparrow\uparrow}=0.5$. On average we obtained a fidelity of $95.5\%$.

As a general observation we note that in both optical experiments \cite{Vermeyden2015,Harris2016} higher fidelities were achieved. However, demonstrating envariance for two qubits on Quantum Experience poses hardly any technical challenge, and more complicated situations including higher dimensional systems can be studied as well. 

\paragraph{Counter example}

Before we move on to higher dimensional systems, however, we briefly illustrate the judicious choice of the initial states and the unitary maps. Envariance is a symmetry of quantum states, $\ket{\psi_\mc{SE}}$, under a a pair of unitary maps, $U_\mc{S}$ and $U_\mc{E}$. This means that if we apply the same map to a different initial state $|\tilde{\psi}_\mc{SE}\rangle$ the map on $\mc{E}$ does no longer act like the inverse of $U_\mc{S}$, or more mathematically $|\tilde{\psi}_\mc{SE}\rangle \neq U_\mc{E} \,U_\mc{S} |\tilde{\psi}_\mc{SE}\rangle$.

\begin{figure}
\centering
\includegraphics[trim={0 1cm 28cm 3cm},clip,width=.9\textwidth]{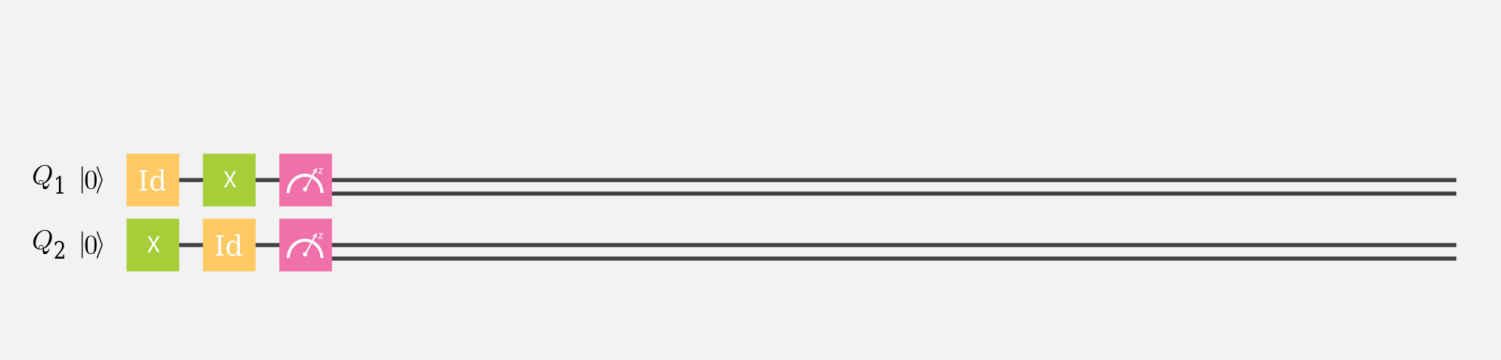}
\caption{\label{figc1} Depiction of a quantum algorithm equivalent, which illustrates the importance of a judicious choice of the initial state. As before, $Q2$ is chosen as system, $\mc{S}$, and $Q1$ is the quantum environment, $\mc{E}$, however we start in $|\tilde{\psi}_\mc{SE}\rangle=\ket{\downarrow\downarrow}$. As in Fig.~\ref{fig1} the unitaries $U_\mc{S}$ and $U_\mc{E}$ are given by $\sigma_x$-gates.}
\end{figure}

To demonstrate the latter comment on Quantum Experience we omitted the initial state preparation, and acted with the two swap operation directly on the default ground state $|\tilde{\psi}_\mc{SE}\rangle=\ket{\downarrow\downarrow}$, cf. Fig.~\ref{figc1}.

Table~\ref{tabc1} summarizes the observed frequencies for several runs. We observe that rather then returning to the initial state $\ket{\downarrow\downarrow}$ the concocted map  $U_\mc{E} \,U_\mc{S}$ produces $\ket{\uparrow\uparrow}$ (to very high probability).
\begin{table}
\centering
\begin{tabular}{c||c|c|c|c}
& $\ket{\downarrow\downarrow}$&$\ket{\downarrow\uparrow}$ &$\ket{\uparrow\downarrow}$ &$\ket{\uparrow\uparrow}$ \\
\hline\hline
theory &0 & 0 &0 &1\\
\hline
run 1 (1024) &0.001 & 0.039  &0.032 &0.982 \\
\hline
run 2 (8192) &0.002 &0.036 &0.032 & 0.931\\
\end{tabular}
\caption{\label{tabc1} Relative frequencies of the final states obtained from the algorithm depicted in Fig.~\ref{figc1} for several separate runs with either 1024 or 8192 shots.}
\end{table}

\subsubsection{Swaps in 3 qubits}
\begin{figure}
\centering
\includegraphics[trim={0 1cm 28cm 1cm},clip,width=.9\textwidth]{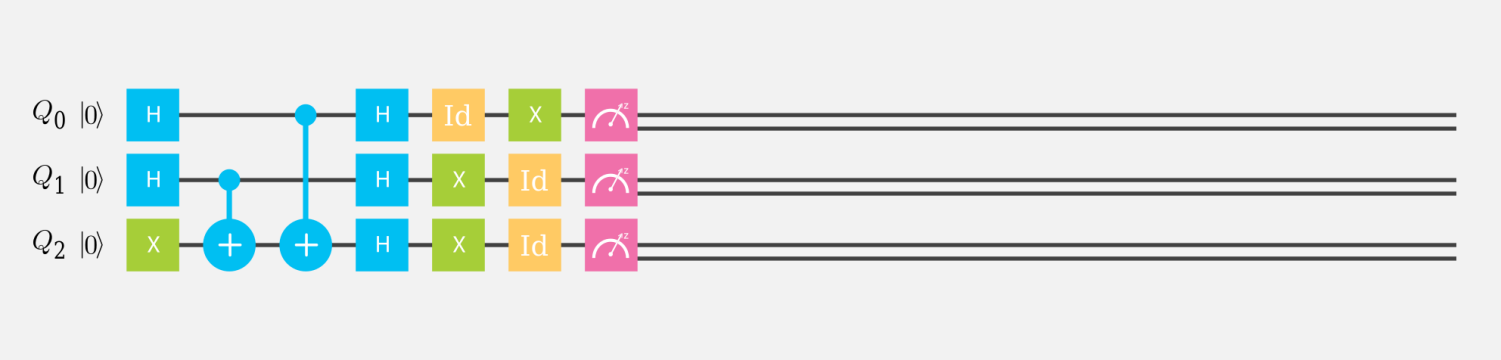}
\caption{\label{fig2} Depiction of a quantum algorithm equivalent to the envariant swap operation, cf. Eqs.~\eqref{eq05}-\eqref{eq06}. $Q2$ and $Q1$ are chosen to span the system, $\mc{S}$, and $Q1$ is the quantum environment, $\mc{E}$. Our desired initial state, $\ket{\psi_\mc{SE}}\propto\ket{\uparrow\uparrow\uparrow}+\ket{\downarrow\downarrow\downarrow}$, is obtained by performing a Hadamard-gates, $H$ (blue box) on $Q1$ and $Q2$, and an $\sigma_X$-gate (green box) on $Q2$, followed by two CNOTs entangling $\mc{S}$ and $\mc{E}$, and concluded by final Hadamard-gates on all three qubits. The swap in $\mc{S}$ is realized by two bitflips, i.e., two $\sigma_X$-gates in $\mc{S}$, which can be counteracted by a single bitflip in $\mc{E}$. The pink boxes at the end of the algorithm depict again the projective measurement of the qubits' states.}
\end{figure}

As a second example we consider a situation, in which the system $\mc{S}$ consist of two qubits, while the environment $\mc{E}$ is still given by only a single qubit. In complete analogy to the previous example, we start with the GHZ-state, $\ket{\psi_\mc{SE}}\propto\ket{\uparrow\uparrow}_\mc{S}\otimes\ket{\uparrow}_\mc{E}+\ket{\downarrow\downarrow}_\mc{S}\otimes\ket{\downarrow}_\mc{E}\equiv\ket{\uparrow\uparrow\uparrow}+\ket{\downarrow\downarrow\downarrow}$. As before, we will now demonstrate that such a state is envariant under a swap operation, namely 
\begin{equation}
\label{eq05}
 \begin{tikzpicture}[>=stealth,baseline,anchor=base,inner sep=0pt]
      \matrix (foil) [matrix of math nodes,nodes={minimum height=0.5em}] {
         & {\color{blue}\ket{\uparrow\uparrow}_\mc{S}} & \otimes & \ket{\uparrow}_\mc{E} &  & + &  & {\color{blue}\ket{\downarrow\downarrow}_\mc{S}} & \otimes & \ket{\downarrow}_\mc{E} &  \xrightarrow{\quad {\color{blue}U_\mc{S}}\quad} 
{\color{blue}\ket{\downarrow\downarrow}_\mc{S}}\otimes\ket{\uparrow}_\mc{E}+{\color{blue}\ket{\uparrow\uparrow}_\mc{S}}\otimes\ket{\downarrow}_\mc{E}\,.\\
      };
      \path[->] ($(foil-1-2.north)+(0,1ex)$)   edge[blue,bend left=45]    ($(foil-1-8.north)+(0,1ex)$);
      \path[<-] ($(foil-1-2.south)-(0,1ex)$)   edge[blue,bend left=-45]    ($(foil-1-8.south)-(0,1ex)$);
    \end{tikzpicture}
\end{equation}
which can be restored by
  \begin{equation}
\label{eq06}
    \begin{tikzpicture}[>=stealth,baseline,anchor=base,inner sep=0pt]
      \matrix (foil) [matrix of math nodes,nodes={minimum height=0.5em}] {
         & \ket{\downarrow\downarrow}_\mc{S} & \otimes & {\color{red}\ket{\uparrow}_\mc{E}} &  & + &  & \ket{\uparrow\uparrow}_\mc{S} & \otimes & {\color{red}\ket{\downarrow}_\mc{E}} & \xrightarrow{\quad {\color{red}U_\mc{E}}\quad} 
\ket{\downarrow\downarrow}_\mc{S}\otimes{\color{red}\ket{\downarrow}_\mc{E}}+\ket{\uparrow\uparrow}_\mc{S}\otimes{\color{red}\ket{\uparrow}_\mc{E}}\,. \\
      };
      \path[->]  ($(foil-1-4.north)+(0,1ex)$) edge[red,bend left=45] ($(foil-1-10.north)+(0,1ex)$);
       \path[<-]  ($(foil-1-4.south)-(0,1ex)$) edge[red,bend left=-45] ($(foil-1-10.south)-(0,1ex)$);
    \end{tikzpicture}
  \end{equation}
The whole algorithm is depicted in Fig.~\ref{fig2}.

Table.~\ref{tab2} summarizes the outcome of this experiment. We observe that while envariance is still evidently demonstrated the initial state is only restored with an average fidelity of 88.7$\%$. 

The lower fidelity is readily understood as a consequence of decoherence, which is more effective in the more complicated quantum system.
\begin{table}
\centering
\begin{tabular}{c||c|c|c|c|c|c|c|c||c}
& $\ket{\downarrow\downarrow\downarrow}$&$\ket{\downarrow\downarrow\uparrow}$ & $\ket{\downarrow\uparrow\downarrow}$ & $\ket{\uparrow\downarrow\downarrow}$ & $\ket{\downarrow\uparrow\uparrow}$ &$\ket{\uparrow\downarrow\uparrow}$ &$\ket{\uparrow\uparrow\downarrow}$ &$\ket{\uparrow\uparrow\uparrow}$&$B$ \\
\hline\hline
theory &0.5 & 0 &0&0&0&0&0 &0.5&1\\
\hline
run 1 (8192) &0.420 & 0.029 &0.033 &0.056&0.047 &0.062 &0.036 &0.316 &0.856\\
\hline
run 2 (1024) &0.427 & 0.032&0.016 &0.035&0.038  &0.073 &0.021 &0.357 &0.885\\
\hline
run 3 (8192) & 0.483&0.021 &0.028  &0.016 & 0.026 &0.040 &0.022 &0.365 &0.919
\end{tabular}
\caption{\label{tab2} Relative frequencies of the final states obtained from the algorithm depicted in Fig.~\ref{fig2} for several separate runs with either 1024 or 8192 shots;  $B$  is the classical fidelity \eqref{eq04} with respect to the theoretically expected values.}
\end{table}

\subsubsection{Swaps in 5 qubits}
\begin{figure}
\centering
\includegraphics[trim={0 1cm 16cm 1cm},clip,width=.9\textwidth]{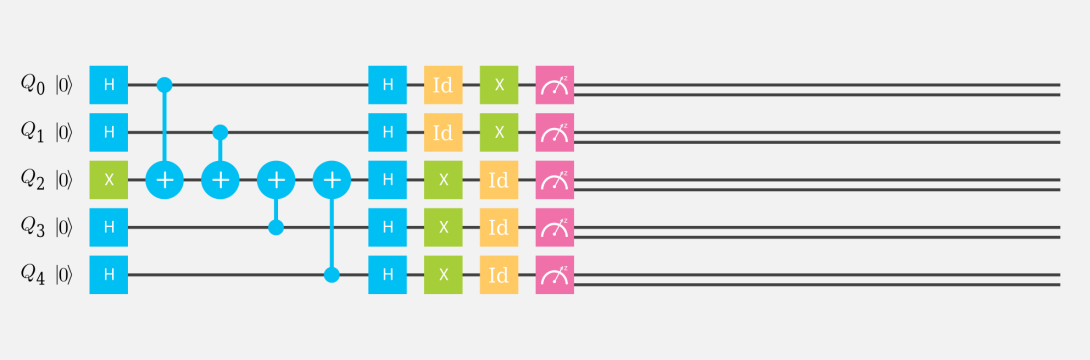}
\caption{\label{fig3}Depiction of a quantum algorithm equivalent to the envariant swap operation, cf. Eqs.~\eqref{eq07}-\eqref{eq08}. $Q4$, $Q3$ and $Q2$ are chosen to span the system, $\mc{S}$, and $Q1$ and $Q0$ are the quantum environment, $\mc{E}$. The initial GHZ-state, $\ket{\psi_\mc{SE}}\propto\ket{\uparrow\uparrow\uparrow\uparrow\uparrow}+\ket{\downarrow\downarrow\downarrow\downarrow\downarrow}$, is obtained by performing Hadamard-gates on $Q4$, $Q3$, $Q1$ and $Q0$, and a $\sigma_X$-gate on $Q2$, followed by CNOTs entangling $\mc{S}$ and $\mc{E}$, and concluded by final Hadamard-gates on all five qubits. The swap in $\mc{S}$ is realized by three bitflips, i.e., three $\sigma_X$-gates in $\mc{S}$, which can be counteracted by two bitflips in $\mc{E}$.}
\end{figure}

The largest ``quantum universe'' that can be realized on Quantum Experience consists of 5 qubits. Hence, we chose as a final example for enavariant swaps the quantum system $\mc{S}$ to comprise 3 qubits, which leaves 2 qubits for $\mc{E}$. Accordingly, we have,
\begin{equation}
\label{eq07}
 \begin{tikzpicture}[>=stealth,baseline,anchor=base,inner sep=0pt]
      \matrix (foil) [matrix of math nodes,nodes={minimum height=0.5em}] {
         & {\color{blue}\ket{\uparrow\uparrow\uparrow}_\mc{S}} & \otimes & \ket{\uparrow\uparrow}_\mc{E} &  & + &  & {\color{blue}\ket{\downarrow\downarrow\downarrow}_\mc{S}} & \otimes & \ket{\downarrow\downarrow}_\mc{E} &  \xrightarrow{\quad {\color{blue}U_\mc{S}}\quad} 
{\color{blue}\ket{\downarrow\downarrow\downarrow}_\mc{S}}\otimes\ket{\uparrow\uparrow}_\mc{E}+{\color{blue}\ket{\uparrow\uparrow\uparrow}_\mc{S}}\otimes\ket{\downarrow\downarrow}_\mc{E}\,.\\
      };
      \path[->] ($(foil-1-2.north)+(0,1ex)$)   edge[blue,bend left=45]    ($(foil-1-8.north)+(0,1ex)$);
      \path[<-] ($(foil-1-2.south)-(0,1ex)$)   edge[blue,bend left=-45]    ($(foil-1-8.south)-(0,1ex)$);
    \end{tikzpicture}
\end{equation}
which can be restored by
  \begin{equation}
\label{eq08}
    \begin{tikzpicture}[>=stealth,baseline,anchor=base,inner sep=0pt]
      \matrix (foil) [matrix of math nodes,nodes={minimum height=0.5em}] {
         & \ket{\downarrow\downarrow\downarrow}_\mc{S} & \otimes & {\color{red}\ket{\uparrow\uparrow}_\mc{E}} &  & + &  & \ket{\uparrow\uparrow\uparrow}_\mc{S} & \otimes & {\color{red}\ket{\downarrow\downarrow}_\mc{E}} & \xrightarrow{\quad {\color{red}U_\mc{E}}\quad} 
\ket{\downarrow\downarrow\downarrow}_\mc{S}\otimes{\color{red}\ket{\downarrow\downarrow}_\mc{E}}+\ket{\uparrow\uparrow\uparrow}_\mc{S}\otimes{\color{red}\ket{\uparrow\uparrow}_\mc{E}}\,. \\
      };
      \path[->]  ($(foil-1-4.north)+(0,1ex)$) edge[red,bend left=45] ($(foil-1-10.north)+(0,1ex)$);
       \path[<-]  ($(foil-1-4.south)-(0,1ex)$) edge[red,bend left=-45] ($(foil-1-10.south)-(0,1ex)$);
    \end{tikzpicture}
  \end{equation}
The corresponding algorithm is depicted in Fig.~\ref{fig3}.

The possible outcome for such an algorithm includes $2^5=32$ states, and therefore we only list the frequencies with which we obtained the theoretically expected states in Tab.~\ref{tab3}. On average we only found the resorted state with a fidelity of 73.6$\%$, and hence the demonstration is not quite as convincing as for the above two examples with smaller Hilbert spaces. 

Our best guess is that the bigger aberrations originate in decoherence and non-perfect implementation of the single gates. Further analysis of these results, however, without direct access to and benchmarking of the experimental systems is hardly feasible.
\begin{table}
\centering
\begin{tabular}{c||c|c|c||c}
& $\ket{\downarrow\downarrow\downarrow\downarrow\downarrow}$&other&$\ket{\uparrow\uparrow\uparrow\uparrow\uparrow}$&$B$ \\
\hline\hline
theory &0.5 & 0 &0.5&1\\
\hline
run 1 (8192) &0.297 & 0.476&0.227&0.722\\
\hline
run 2 (1024) & 0.273& 0.501&0.227&0.706\\
\hline
run 3 (8192) & 0.308&0.470 &0.222&0.726\\
\hline
run 4 (8192) &0.348 &0.376 &0.276 &0.789
\end{tabular}
\caption{\label{tab3} Relative frequencies of the final states obtained from the algorithm depicted in Fig.~\ref{fig3} for several separate runs with either 1024 or 8192 shots;  $B$  is the classical fidelity \eqref{eq04} with respect to the theoretically expected values.}
\end{table}

\subsection{Other operations on 2 qubits}

To conclude the analysis with two further examples we now return to  the smallest ``quantum universe'', in which $\mc{S}$ and $\mc{E}$ are given by single qubits. The next two examples elucidate that $\ket{\psi_\mc{SE}}\propto\ket{\uparrow}_\mc{S}\otimes\ket{\uparrow}_\mc{E}+\ket{\downarrow}_\mc{S}\otimes\ket{\downarrow}_\mc{E}$ is not only envariant under swaps, cf. Eqs.~\eqref{eq04}-\eqref{eq05}, but actually under all unitary maps \cite{Deffner2016}.

\subsubsection{Creating and destroying superpositions}

\begin{figure}
\centering
\includegraphics[trim={0 1cm 28cm 3cm},clip,width=.9\textwidth]{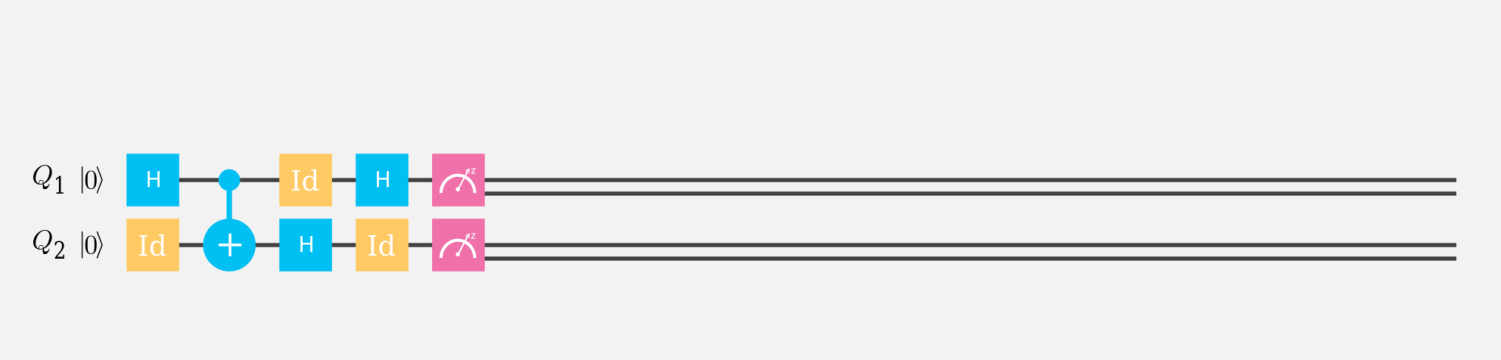}
\caption{\label{fig4}Depiction of a quantum algorithm equivalent to an envariant operation creating and destroying superpositions. $Q2$ is chosen as system, $\mc{S}$, and $Q1$ is the quantum environment, $\mc{E}$. As before, our desired initial state, $\ket{\psi_\mc{SE}}\propto\ket{\uparrow\uparrow}+\ket{\downarrow\downarrow}$, is obtained by performing a Hadamard-gate, $H$ on $Q1$, followed by a CNOT.}
\end{figure}

As a first example consider the algorithm of Fig.~\ref{fig4}. In comparison to the previous example in Fig.~\ref{fig1} we replaced the bitflip operation by Hadamard-gates. Physically, Hadamard-gates can be understood as operations that create and destroy superpositions. One easily convinces oneself that $\ket{\psi_\mc{SE}}\propto\ket{\uparrow}_\mc{S}\otimes\ket{\uparrow}_\mc{E}+\ket{\downarrow}_\mc{S}\otimes\ket{\downarrow}_\mc{E}$ is also envariant under $U_\mc{S}=H_\mc{S}\otimes\id_\mc{E}$, where $H=1/2\,(\ket{\downarrow}\bra{\downarrow}+\ket{\uparrow}\bra{\downarrow}+\ket{\downarrow}\bra{\uparrow}-\ket{\uparrow}\bra{\uparrow})$ and $U_\mc{E}=\id_\mc{S}\otimes H_\mc{E}$.

In Tab.~\ref{tab4} we summarize our findings. We observe that once again envariance is demonstrated with an average fidelity of 95.8$\%$.
\begin{table}
\centering
\begin{tabular}{c||c|c|c|c||c}
& $\ket{\downarrow\downarrow}$&$\ket{\downarrow\uparrow}$ &$\ket{\uparrow\downarrow}$ &$\ket{\uparrow\uparrow}$&$B$ \\
\hline\hline
theory &0.5 & 0 &0 &0.5&1\\
\hline
run 1 (1024) &0.515 & 0.036 &0.040 &0.409&0.960\\
\hline
run 2 (8192) & 0.518& 0.036&0.048 &0.398&0.955\\
\end{tabular}
\caption{\label{tab4} Relative frequencies of the final states obtained from the algorithm depicted in Fig.~\ref{fig4} for several separate runs with either 1024 or 8192 shots;  $B$  is the classical fidelity \eqref{eq04} with respect to the theoretically expected values.}
\end{table}

\subsubsection{Superpositions and bitflips}
\begin{figure}
\centering
\includegraphics[trim={0 1cm 28cm 3cm},clip,width=.9\textwidth]{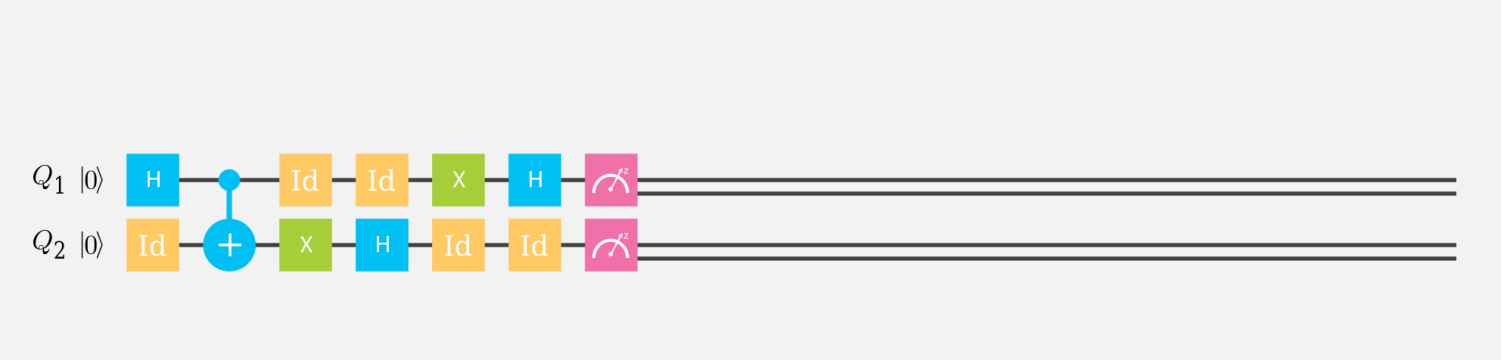}
\caption{\label{fig5}Depiction of a quantum algorithm equivalent to an envariant operation consisting of creating and destroying superpositions, and bitflips. $Q2$ is chosen as system, $\mc{S}$, and $Q1$ is the quantum environment, $\mc{E}$. In contrast to previous examples, here we chose the unitary map on $\mc{S}$ to be given not only by a single gate, but rather as a concoction of a Hadamard-gate and a $\sigma_X$-gate, i.e., $U_\mc{S}=H_\mc{S}\cdot\sigma_{X_\mc{S}}\otimes \id_\mc{E}$. }
\end{figure}

Finally, we chose $U_\mc{S}$ to be a concoction of a Hadamard-gate and a $\sigma_X$-gate, i.e., $U_\mc{S}=H_\mc{S}\cdot\sigma_{X_\mc{S}}\otimes \id_\mc{E}$. The corresponding algorithm is illustrated in Fig.~\ref{fig5}.

Table~\ref{tab5} summarizes our findings. Envariance is observed with an average fidelity of  95.8$\%$.
\begin{table}
\centering
\begin{tabular}{c||c|c|c|c||c}
& $\ket{\downarrow\downarrow}$&$\ket{\downarrow\uparrow}$ &$\ket{\uparrow\downarrow}$ &$\ket{\uparrow\uparrow}$&$B$ \\
\hline\hline
theory &0.5 & 0 &0 &0.5&1\\
\hline
run 1 (8192) &0.520 & 0.057 &0.036 &0.387&0.950\\
\hline
run 2 (8192) & 0.552& 0.031&0.029 &0.387&0.965\\
\end{tabular}
\caption{\label{tab5} Relative frequencies of the final states obtained from the algorithm depicted in Fig.~\ref{fig5} for several separate runs with either 1024 or 8192 shots;  $B$  is the classical fidelity \eqref{eq04} with respect to the theoretically expected values.}
\end{table}

\section{\label{con}Concluding remarks}

In the present analysis we have described six pedagogically chosen experiments on IBM's Quantum Experience to elucidate a purely quantum symmetry. We have seen that demonstrating the concepts does not pose any technical challenges, and that these experiments can be reproduced with little effort. However, we have also seen that quality of the experimental outcome still lacks behind technically more challenging experiments in quantum optics. Vermeyden \etal \cite{Vermeyden2015} achieved Bhattacharya coefficients of $(99.963\pm 0.005)\%$ and Harris \etal \cite{Harris2016} $(99.15\pm 0.41)\%$, whereas our experiments did not perform better than $96\%$. However, Quantum Experience has the advantage that it is very simple to operate and that principally ``quantum universes'' of up to five qubits are available. Our experiment with five qubits revealed that  decoherence and (possibly) non-perfect implementation of the quantum gate are significant, and that even for the simplest possible example evariance could not be demonstrated convincingly.

\ack{The author acknowledges support by the U.S. National Science Foundation under Grant No. CHE-1648973 and the use of IBM's Quantum Experience for this work. The views expressed are those of the author and do not reflect the official policy or position of IBM or the IBM Quantum Experience team.}

\section*{References}

\bibliographystyle{unsrt}
\bibliography{references_env_IBM}

\end{document}